\documentclass[prd,twocolumn,nofootinbib,floatfix,
superscriptaddress]{revtex4}
 \pdfoutput=1
 
\usepackage{graphicx}
\usepackage{epsfig}
\usepackage{bm}
\usepackage[T1]{fontenc}
\usepackage[latin9]{inputenc}
\usepackage{amssymb}
\usepackage{float}
\usepackage{amsmath}
\usepackage{dcolumn}
\usepackage{cancel}
\usepackage[colorlinks]{hyperref}
\usepackage[usenames,dvipsnames]{color}
\hypersetup{
     breaklinks=true,
    pdfstartview={FitH},    
    colorlinks=true,       
    linkcolor=blue,          
    citecolor=red,        
    filecolor=magenta,      
    urlcolor=blue,           
    anchorcolor=green,      
    linktocpage=true
}

\newcommand{\be}{\begin{equation}}
\newcommand{\ee}{\end{equation}}

\begin{document}
\title{Baryogenesis in non-extensive Tsallis Cosmology}

\author{Giuseppe Gaetano Luciano}
\email{gluciano@sa.infn.it}
\affiliation{Dipartimento di Fisica, Universit\`a di Salerno, Via Giovanni Paolo 
II, 132 I-84084 Fisciano (SA), Italy}
\affiliation{ INFN, Sezione di Napoli, Gruppo collegato di Salerno, Via Giovanni Paolo II, 132 I-84084 Fisciano (SA), Italy}
\affiliation{Departament de Matem\`atica, Universitat de Lleida, Av. Jaume II, 69, 25001, Lleida, Catalonia, Spain}

\author{Jaume Gin\'e}
\email{gine@matematica.udl.cat}
\affiliation{Departament de Matem\`atica, Universitat de Lleida, Av. Jaume II, 69, 25001, Lleida, Catalonia, Spain}

\begin{abstract}
Non-extensive Tsallis thermostatistics is a widespread
paradigm to describe large-scale gravitational systems.
In this work we use Tsallis Cosmology 
to study thermodynamic gravity and derive
modified Friedmann equations. We show that corrections induced
by non-extensivity affect the Hubble function
evolution during the radiation-dominated epoch. In turn, this leads
to non-trivial modifications of the mass density and pressure content of the Universe, which provide a viable 
mechanism allowing for baryogenesis, even in the presence of the standard interaction between the Ricci scalar and baryon current.
By demanding consistency with current observational bounds on baryogenesis, we constrain Tsallis $\delta$ parameter to be $|\delta -1 |\simeq10^{-3}$. 
Based on the recently established connection between Tsallis thermostatistics 
and the quantum gravitational generalization of the uncertainty principle at Planck scale (GUP),
we finally show that this bound is in agreement with the estimation of the GUP parameter predicted by many quantum gravity models. 
\end{abstract}


\maketitle

\section{Introduction}
\label{Intro}

There are indications from various scenarios and results, 
such as black hole mechanics~\cite{Bard}, Bekenstein-Hawking (BH) formula~\cite{Beke,Hawk,ScardBek},  holographic principle~\cite{tHooft,Sussk} and 
the recent firewall puzzle~\cite{Almhe, Braunst}, that the concepts
of \emph{gravity} and \emph{entropy} are non-trivially connected, 
possibly intertwined in the would-be theory of quantum gravity. 
This appears even more evident in the gravity-thermodynamics conjecture
by Jacobson~\cite{Jaco}, who showed that Einstein's equations
describing relativistic gravitation can be naturally derived by 
assuming that the Bekenstein bound and the laws of thermodynamics hold true.
The ensuing research framework is often referred to as \emph{thermodynamic gravity}. 

Thermodynamic gravity has
been largely addressed over the years~\cite{Pad,Verlinde}. One of its most fascinating implications is the possibility to deduce the
cosmological Friedmann equations from the first law of thermodynamics on the apparent horizon
of a Friedmann-Robertson-Walker (FRW) Universe (see, for instance,~\cite{Wang,Cai,Eling,Akbar1,Akbar1b,Frolov,Akbar2,Sheykhi}). 
This connection has also been studied in various models of modified gravity, 
such as Lanczos-Lovelock~\cite{Para}, Horava-Lifshitz~\cite{Jamil}, quadratically extended~\cite{Fan}, Gauss-Bonnet and running vacuum~\cite{MavSar} models,  
highlighting the general validity of the gravity-thermodynamics conjecture.

The above analysis has been developed 
within the framework of standard thermodynamics based
on Boltzmann-Gibbs entropy. However, for systems 
with divergent partition function such as large-scale gravitational systems, 
Boltzmann-Gibbs theory is not applicable, requiring a non-additive generalization of the entropy definition. Among various proposals, 
Tsallis prescription~\cite{Tsallis,Tsallisbo} has been providing encouraging results
in the description of many complex (strongly correlated) systems, \emph{e.g.}, self-gravitating stellar systems~\cite{App1,App3}, black holes~\cite{Tsallis3}, background radiation~\cite{App7}, neutrinos~\cite{Kanane,Lucianomix,Lucianomix2}, holographic dark energy~\cite{Saridakis:2018unr} and dark matter~\cite{Guha}, low-dimensional dissipative systems~\cite{Tsallis4} and polymer chains~\cite{Polch}. 

On the basis of statistical arguments, Tsallis and Cirto~\cite{Tsallis3} 
have argued that
the entropy of a black hole might not obey the traditional BH
 area-law, but rather
\begin{equation}
\label{ModAreaLaw}
S_\delta=\gamma {\left(\frac{A}{\ell_p^{2}}\right)}^\delta\,,
\end{equation}
where $A$ is the black hole horizon area, $\gamma>0$ 
a constant factor and $\ell_p$ the Planck length\footnote{Throughout the whole manuscript, we work in natural units $\hbar=c=k_B=1$.}.
Deviation from extensivity is quantified by the Tsallis exponent
$\delta>0$, which gives rise to a probability distribution decaying
asymptotically as a power law (rather than an exponential). 
It must be emphasized that this parameter is not
fixed by the theory, the only expectation being $|\delta-1|\ll1$.
Clearly, the standard BH entropy 
is recovered for $\delta\rightarrow1$, provided that $\gamma\rightarrow1/4$ in this limit. 
Although not contemplated in the original formulation by Tsallis, 
the possibility of a running  $\delta$ has been
considered in Ref.~\cite{App13} and~\cite{Lucianomix,Lucianomix2} for 
quantum gravity and field theoretical systems, respectively. 

In recent years, Tsallis entropy
has aroused a lot of interest in cosmological contexts.
For instance, in Ref.~\cite{App14} 
it has been shown that non-extensivity translates into 
a modification of the gravitational constant and, accordingly, 
of the dark content of the Universe. 
Likewise, by assuming Eq.~\eqref{ModAreaLaw} 
as entropy law for the Universe apparent horizon, 
one can derive modified Friedmann equations and
a Tsallis driven cosmic evolution~\cite{Lymp,She3,Noji}
for any spatial curvature~\cite{She2}.
Interestingly enough, for a particular choice of Tsallis parameter  
the model of~\cite{She2} is able to reproduce the
cosmic acceleration without invoking  dark energy. A quite similar
result has been exhibited in Ref.~\cite{She3}, 
showing that the
modified Newton equation on large scales can explain the
asymptotic flatness of 
galaxy rotation curves without the need of dark matter, 
provided that $\delta\lesssim1/2$. 

While being intensively investigated for late-time Cosmology, comparably less attention has been devoted to the study of Tsallis thermodynamics in the early Universe.
One of the most debated problems in this context
is the observed baryon (\emph{i.e.} matter/anti-matter) asymmetry generated in the radiation dominated era. 
For this asymmetry to occur, it is known that three (Sakharov) conditions are to be satisfied~\cite{Saka}, at least  
in the simplest versions of baryogenesis:
\emph{i)} baryon number $B$ violation,  
\emph{ii)} $C$-symmetry and $CP$-symmetry violation and
\emph{iii)} out-of-thermal-equilibrium interactions. 
In the standard cosmological model based on Boltzmann-Gibbs entropy, the predicted baryon asymmetry equals zero, due to the last condition
not being satisfied\footnote{It should be said that in some extended models, such as Grand Unified Theories (GUTs)~\cite{Buras1,Georgi,Dimopoulos,Langacker,Kolb}, all three of Sakharov's conditions are naturally satisfied. However, in this work we aim at explaining baryogenesis by following a different approach built on a modification of the entropy area-law.}. 
Therefore, despite strong evidences from Cosmic Microwave Background~\cite{CMB} and Big Bang Nucleosynthesis~\cite{BBN} measurements, the origin of this phenomenon is not yet understood, leaving room for disparate explanations~\cite{Kugo, Cohen, Davoudiasl,Canetti, Alex,Lamb}. 

Starting from the above premises,  in this work
we investigate baryogenesis
in Tsallis Cosmology.
In particular, we adopt the entropy~\eqref{ModAreaLaw} for the horizon degrees of freedom of the FRW Universe and derive modified Friedmann equations. 
This leads to corrected mass density and pressure content of the Universe, which provide an effective mechanism to break the thermal equilibrium in the radiation dominated era. 
Along with the conventional 
coupling between spacetime and baryon current 
satisfying the first two Sakharov conditions, 
this scenario ensures that all Sakharov's criteria are met, 
thus allowing for baryogenesis. 
We exploit this result to constrain Tsallis Cosmology 
by comparison with observational bounds on baryon asymmetry.

It is worth noting that the rationale behind our analysis takes its cue from situations typically occurring in low-energy systems, such as crystalline membranes, where thermal fluctuations are intimately connected with long-range dipole interactions~\cite{Membr}, or strong electrolytes, which exhibit long-distance behaviors when driven out of equilibrium by external fields~\cite{Elect}. Here, 
we are somehow reversing the perspective, exploring the possibility
of a deviation from thermality induced by long-range correlations
in gravitational systems obeying Tsallis description. In a broader sense, this framework
is motivated by the fact that Eq.~\eqref{ModAreaLaw} works well for macroscopic states that exhibit
some kind of (quasi-)stationarity or metastability and 
for driven non-equilibrium systems with
large-scale temperature fluctuations (superstatistical systems~\cite{Beck,Wilk,Beckbis}). 

In passing, we mention that a similar study has been developed
in Ref.~\cite{Lamb}. In that case the mechanism responsible for 
baryon asymmetry is identified with quantum gravity corrections
embedded in a generalized uncertainty principle (GUP) at Planck scale.  
By exploiting the recently established connection between 
Tsallis statistics and GUP~\cite{JizbaLuc}, we show that
our bound on $\delta$ is indeed consistent with 
the value of the GUP deformation parameter predicted by many quantum gravity models.

The layout of the paper is as follows. In the next Section 
we derive the modified Friedmann equations in Tsallis Cosmology.
Section~\ref{CTC} is devoted to the investigation of baryon asymmetry problem.
Consistency between non-extensive Tsallis statistics and 
Planck-scale deformed uncertainty relations is then discussed in Section~\ref{CBT}.
Conclusion and outlook are summarized in Section~\ref{DC}.

\section{Modified Friedmann equations in Tsallis Cosmology}
\label{MFE}
In this Section we derive the modified Friedmann equations
in FRW Universe within the framework of Tsallis Cosmology. 
Suppose the spacetime geometry is given by the $(1+3)$-dimensional metric
\begin{equation}
ds^2=h_{bc}\hspace{0.2mm}dx^{b}dx^{c}+\tilde r^2(d\theta^2+\sin^2\theta d\phi^2)\,,\,\,\,\, b,c=\{0,1\},
\end{equation}
where $h_{bc}=\mathrm{diag}(-1,a^2/(1-kr^2))$ is the metric
of a $(1+1)$-dimensional subspace, $x^b=(t,r)$,
$\tilde r=a(t)r$, with $a(t)$ being the scale factor, $r$ is the comoving
radius and $k$ the (constant) spatial curvature. As shown in~\cite{Akbar2}, the dynamical  apparent horizon is fixed
by the condition $h^{ab}\partial_a\tilde r \partial_b \tilde r=0$, 
which gives for the FRW Universe
\begin{equation}
\label{rtilde}
\tilde r_A=\frac{1}{\sqrt{H^2+{k}/{a^2}}}\,,
\end{equation}
where $H=\dot a(t)/a(t)$ is the Hubble parameter, the dot indicating a time derivative. 

The temperature $T=\kappa/2\pi$ 
associated to such horizon can be derived
from the definition of the surface gravity~\cite{Akbar2}
\begin{equation}
\kappa\,=\,\frac{1}{2\sqrt{-h}}\partial_a\left(\sqrt{-h}h^{ab}\partial_b\tilde r\right)=-\frac{1}{\tilde r_A}\left(1-\frac{\dot{\tilde r}_A}{2H\tilde r_A}\right),
\end{equation}
which gives
\begin{equation}
\label{T}
T\,=\,-\frac{1}{2\pi \tilde r_A}\left(1-\frac{\dot{\tilde r}_A}{2H\tilde r_A}\right). 
\end{equation}

The matter/energy content of the Universe is
described by a perfect fluid. Denoting by $\rho$ and $p$ its mass density
and pressure at equilibrium, respectively, the corresponding 
energy-momentum tensor is 
\begin{equation}
T_{\mu\nu}=(\rho+p)u_\mu u_{\nu}+pg_{\mu\nu}\,,
\end{equation} 
where $u_{\mu}$ is the four-velocity of the fluid. In turn, the conservation equation
$\nabla_{\mu}T^{\mu\nu}=0$ for the FRW Universe implies
the continuity equation 
\begin{equation}
\label{ce}
\dot \rho=-3H(\rho+p)\,.
\end{equation}

Now, Friedmann equations in the bulk of the Universe
are obtained by considering the first law of thermodynamics
on the apparent horizon 
\begin{equation}
\label{FLT}
dE=TdS+WdV\,,
\end{equation}
where $E=\rho V$ is the total energy content of the Universe
of $3$-dimensional spherical volume $V=4\pi \tilde r_A^3/3$
and horizon surface area $A=4\pi\tilde r_A^2$. $W$ is the
work density done by the volume
change of the Universe, 
\begin{equation}
\label{W}
W=-\frac{1}{2}T^{bc}h_{bc}=\frac{1}{2}(\rho-p)\,. 
\end{equation}

A comment is now in order: in Tsallis model it has been argued that
non-extensivity translates into a non-trivial
modification of both core thermodynamic relations and   
standard definitions of pressure and temperature (see, for instance, \cite{Pennini}). However, following the analysis of~\cite{Sheykhi,She3,She2,Noji,Lamb,goshal}, such corrections can be neglected as a first approximation in cosmological contexts.
Having this in mind, we henceforth adopt the usual
definitions for $p$ and $T$, 
as well as the first law of thermodynamics in its
traditional form~\eqref{FLT}.

Plugging Eqs.~\eqref{T}, \eqref{ce} and \eqref{W}
into~\eqref{FLT}, after simple algebra one obtains the first modified
Friedmann equation
\begin{equation}
\label{fFE}
-4\pi G \left(\rho+p\right)=\left(\dot H-\frac{k}{a^2}\right)f'(A),
\end{equation}
where $f(A)$ is defined in such a way as
\begin{equation}
\label{Sbis}
S=\frac{f(A)}{4G}\,\,\Longrightarrow\,\, \frac{dS}{dA}=\frac{f'(A)}{4G}\,.
\end{equation}
Notice that $G=\ell_p^2=1/m_p^2$ in our units convention, where
$m_p$ is the Planck mass. Similarly, by use of Eq.~\eqref{ce}, 
the second Friedmann equation reads
\begin{equation}
\label{SecEq}
\frac{8\pi G}{3}\rho=-4\pi\int \frac{f'(A)}{A^2}dA\,.
\end{equation}
We remark that at this stage 
we have not yet used Tsallis definition for the horizon entropy. 
Thus, Eqs.~\eqref{fFE} and~\eqref{SecEq} 
hold true for any modified entropy obeying Eq.~\eqref{Sbis}.
Clearly, by comparison with Eq.~\eqref{ModAreaLaw}, in Tsallis cosmological model we obtain
\begin{equation}
\label{fA}
f'(A)\equiv f'_\delta(A)=4\gamma\hspace{0.2mm} \delta \left(\frac{A}{G}\right)^{\delta-1}\,. 
\end{equation}
In the Boltzmann-Gibbs limit of $\delta\rightarrow1$, we have $f'_\delta(A)\rightarrow1$, consistently with the standard BH area law
$S=A/(4G)$. This allows us to straightforwardly recover
the usual Friedmann equations.

By use of Eq.~\eqref{fA}, the modified Friedmann equations~\eqref{fFE} and~\eqref{SecEq} take the form
\begin{eqnarray}
\label{Quatpig}
-4\pi G\left(\rho+p\right)&=&4\gamma\hspace{0.2mm}\delta\left(\dot H-\frac{k}{a^2}\right)\left(\frac{A}{G}\right)^{\delta-1}\,,\\[2mm]
\label{SFE}
\frac{8\pi G}{3}\rho&=&\frac{16\pi\hspace{0.2mm}\gamma\hspace{0.2mm}\delta}{(2-\delta)}\frac{A^{\delta-2}}{G^{\delta-1}}\,+\,c\,,
\end{eqnarray}
respectively. The integration constant $c$ 
can be fixed by imposing the boundary condition
$\rho\equiv\rho_{\mathrm{vac}}\rightarrow\Lambda$ for $A\rightarrow\infty$ in the vacuum energy dominated era, $\Lambda$ being the cosmological constant. This gives 
\begin{equation}
c=\frac{8\pi G}{3}\Lambda\,,
\end{equation} 
where we have taken into account
that $|\delta-1|\ll1$ (see below Eq.~\eqref{ModAreaLaw}). 
We then obtain
\begin{equation}
\frac{8\pi G}{3}\left(\rho-\Lambda\right)=\frac{16\pi\hspace{0.2mm}\gamma\hspace{0.2mm}\delta}{(2-\delta)}\frac{A^{\delta-2}}{G^{\delta-1}}\,.
\end{equation}

Modified Friedmann equations can be further manipulated
by expressing the horizon surface area as
\begin{equation}
A=4\pi \tilde r_A^2=\frac{4\pi}{H^2+{k}/{a^2}}\,. 
\end{equation} 
By inserting into Eqs.~\eqref{Quatpig} and~\eqref{SFE}, we get
\begin{eqnarray}
\label{firstbisEq}
-4\pi G\left(\rho+p\right)&\hspace{-0.8mm}=\hspace{-0.8mm}&\frac{4^\delta\pi^{\delta-1}\hspace{0.2mm} \gamma\hspace{0.2mm}\delta}{G^{\delta-1}}\left(\dot H-\frac{k}{a^2}\right)\left(H^2+\frac{k}{a^2}\right)^{1-\delta} \\[2mm]
\label{secbisEq}
\frac{8\pi G}{3}\left(\rho-\Lambda\right)&\hspace{-0.8mm}=\hspace{-0.8mm}&\frac{4^\delta \pi^{\delta-1}\hspace{0.2mm} \gamma\hspace{0.2mm}\delta}{\left(2-\delta\right)G^{\delta-1}}\left(H^2+\frac{k}{a^2}\right)^{2-\delta}\,.
\end{eqnarray}

If we now apply the above model to the radiation dominated era, 
we can safely neglect the tiny observed cosmological constant $\Lambda$
and set the spatial curvature constant $k=0$, consistently with the
the observed spatially flat Universe. Accordingly, the modified Friedmann equations
become
\begin{eqnarray}
\label{firsterEq}
-4\pi G\left(\rho+p\right)&=&\frac{4^\delta\pi^{\delta-1}\hspace{0.2mm} \gamma\hspace{0.2mm}\delta}{G^{\delta-1}}\,\dot H\,H^{2(1-\delta)}\,, \\[2mm]
\frac{8\pi G}{3}\rho&=&\frac{4^\delta \pi^{\delta-1}\hspace{0.2mm} \gamma\hspace{0.2mm}\delta}{\left(2-\delta\right)G^{\delta-1}}\,H^{2(2-\delta)}\,.
\label{secterEq}
\end{eqnarray}
It is immediate to check that the limit of $\delta\rightarrow1$  
gives back the standard cosmological relations, as expected.

As we shall see in the next Section, 
Eqs.~\eqref{firsterEq} and~\eqref{secterEq} lie at the heart of the computation of Tsallis-modified mass density and pressure content in the radiation dominated era. The evolution of the Universe in Tsallis Cosmology has been studied in~\cite{Lymp}, showing that generalized
Friedmann equations contain extra terms that constitute an effective dark energy sector quantified by the nonextensive parameter $\delta$. 

\section{Constraints on Tsallis Cosmology from Baryon Asymmetry}
\label{CTC}
Observational evidences indicate that our Universe
is mostly made up of matter, rather than balanced amounts 
of matter and anti-matter as predicted by Quantum and Relativistic theories~\cite{Canetti}. The common view is that this asymmetry is generated dynamically 
as the Universe expands and cools. In Ref.~\cite{Saka}
Sakharov listed three necessary conditions  
for the occurrence of baryogenesis: 
\emph{i)} baryon number $B$ violation,  
which is required to produce an excess of baryons over anti-baryons;
\emph{ii)} $C$-symmetry and $CP$-symmetry violation. The first one
assures that the interactions which produce more baryons than anti-baryons are not counterbalanced by interactions with reverse effects.
$CP$-violation is similarly needed, since otherwise equal numbers of left-handed baryons and right-handed anti-baryons would be produced, as well as equal numbers of left-handed anti-baryons and right-handed baryons;
\emph{iii)} out-of-thermal-equilibrium interactions, 
which prevent $CPT$-symmetry from compensating 
between processes increasing and decreasing $B$. 

To meet the conditions $\emph{ii)}$ and $\emph{iii)}$,  
the conventional approach provides for the introduction 
of interactions that violate $C$- and $CP$-symmetry in vacuum
and a period of non-thermal equilibrium for the Universe. 
Nevertheless, several other possible explanations have been
proposed over the years~\cite{Kugo, Cohen, Davoudiasl,Canetti, Alex,Lamb}.  
For instance, in Ref.~\cite{Davoudiasl} a mechanism 
has been suggested involving a dynamical breaking of $CPT$
in an expanding Universe. The key ingredient is a $CP$-violating
interaction in vacuum between the derivative of the Ricci scalar
curvature $\mathcal{R}$ and the baryon number current $J^{\mu}$
in the form~\cite{Davoudiasl}
\begin{equation}
\label{Jmu}
\frac{1}{M_*^2}\int d^4x\sqrt{-g}\,J^{\mu}\partial_{\mu}\mathcal{R}\,,
\end{equation}
where $M_*$ is the cutoff scale of the effective theory, which is
taken to be of order of the reduced Planck mass $M_*= (8\pi G)^{-1/2}\simeq2.4\times10^{18}\,\mathrm{GeV}$. 

To generate baryon asymmetry from Eq.~\eqref{Jmu}, it is also required that there be some $B$-violating process.  In Ref.~\cite{Davoudiasl}
this is expected to take place while maintaining thermal equilibrium.
In this scheme, it can be shown that the interaction~\eqref{Jmu} gives
opposite sign energy contributions that differ for particles and
antiparticles. Then, a net
baryon asymmetry
\begin{equation}
\label{oeq}
\frac{1}{M_*^2}J^{\mu}\partial_{\mu}\mathcal{R}=\frac{1}{M_*^2}\left(n_B-n_{\bar{B}}\right)\mathcal {\dot R}\,,
\end{equation}
can be generated, where $n_B$ and $n_{\bar{B}}$ denote the 
baryon and anti-baryon number density, respectively. 

As argued in Ref.~\cite{Davoudiasl}, dynamical $CPT$-violation
modifies thermal equilibrium in a similar fashion as 
a chemical potential
\begin{equation}
\label{chempot}
\mu_B=-\mu_{\bar B}=-\frac{\mathcal{\dot R}}{M_*^2}\,.
\end{equation}
Once the temperature drops below the decoupling temperature $T_D$ at which $B$-violation goes out
of equilibrium, the imbalance 
\begin{equation}
\label{asym}
n_B-n_{\bar{B}}=\bigg|\frac{g_b}{6}\mu_B T^2\bigg|
\end{equation}
gets frozen, where $g_b\sim \mathcal{O}(1)$ is the
number of intrinsic degrees of freedom of baryons\footnote{See Ref.~\cite{Davoudiasl} for more details on how baryon asymmetry occurs.}. 
Baryon asymmetry can be then quantified as~\cite{Kolb} 
\begin{equation}
\label{quant}
\frac{\eta}{7}=\frac{n_B-n_{\bar{B}}}{s}=
\bigg|\frac{15\,g_b}{4\pi^2\,g_{*s}}\frac{\mathcal{\dot R}}{M_*^2\, T}\bigg|_{T=T_D}\,,
\end{equation}
where
\begin{equation}
s=\frac{2\pi^2g_{*s}}{45}T^3\,,
\end{equation}
is the standard entropy density and $g_{*s}$ the number of degrees of freedom 
for particles contributing to the entropy of the Universe
in the radiation dominated era.
As noted in Ref.~\cite{Kolb}, $g_{*s}$ is nearly equal to the total
number $g_*$ of degrees of freedom of relativistic Standard Model
particles, \emph{i.e.}, $g_{*s}\approx g_*\simeq106$.

Now, from Eq.~\eqref{quant} it is clear that baryogenesis occurs, provided that $\mathcal{\dot R}\neq0$. 
In the standard cosmological model based on thermodynamic
Boltzmann-Gibbs entropy, $\mathcal{\dot R}=0$ in the radiation dominated
era, due to thermal equilibrium still being satisfied. On the other hand, 
by using Tsallis entropy for long-range correlated gravitational systems, 
departure from thermality could be naturally allowed, 
as it appears from the next calculation. 
We stress that hints toward this scenario are provided
by analogue matter systems, such as crystalline membranes~\cite{Membr}
and electrolytes~\cite{Elect}, where out-of-equilibrium
macroscopic states are characterized by emergent long-distance 
behaviors of microscopic constituents. Particularly, 
in Ref.~\cite{Membr} it has been shown that long-range electrostatic (dipole)
interactions have non-trivial effects on thermal fluctuations of freestanding crystalline membranes. Likewise, 
in Ref.~\cite{Elect} power-law (slowly decaying) correlations
are found to affect the dynamics of 
out-of-equilibrium strong electrolytes, resulting in
a Casimir-like fluctuation-induced force between boundaries
confining ions in the system. The appearance of near-thermal equilibrium states
with Tsallis distribution has also been investigated in Ref.~\cite{Cleymans,HEC}
in high energy collisions.

Following~\cite{Lamb}, we parameterize departure from thermal equilibrium
by mass density and pressure variations, which are indicated
by $\delta\rho$ and $\delta p$, respectively. The total mass density $\rho$
and pressure $p$ then read
\begin{eqnarray}
\label{prho}
\rho&=&\rho_0+\delta\rho\,,\\[2mm]
p&=&p_0+\delta p\,,
\label{prho2}
\end{eqnarray}
where $\rho_0$ and $p_0$ denote the corresponding quantity
at equilibrium, \emph{i.e.}
\begin{eqnarray}
\label{equi}
\rho_0&=&\frac{3H^2}{8\hspace{0.2mm}G\hspace{0.2mm}\pi}\,,\\[2mm]
p_0&=&w\hspace{0mm}\rho_0\,.
\label{equi2}
\end{eqnarray}
The latter relation defines the well-known equation of state parameter at equilibrium. At the radiation dominated
era, we have $w=1/3$. 

Based on the previous considerations, 
we expect that both these mass density and pressure fluctuations depend
on the parameter $\delta$ in such a way that 
$\delta\rho,\delta p\rightarrow0$ as $\delta\rightarrow1$. 
The explicit expressions of such corrections 
can be derived by plugging Eqs.~\eqref{prho} and~\eqref{prho2} into the modified
Friedmann equations~\eqref{firsterEq} and~\eqref{secterEq}. 
Specifically, from Eq.~\eqref{secterEq} we obtain
\begin{equation}
\label{deltarho}
\delta\rho=\frac{1}{3}\rho_0\left(-3+\frac{2^{3-\delta}\times3^\delta\,\gamma\hspace{0.3mm}\delta}{\left(2-\delta\right)G^{2(\delta-1)}}\,\rho_0^{1-\delta}
\right).
\end{equation}
Requiring $\delta \rho\rightarrow0$ for $\delta\rightarrow1$ fixes definitively the 
constant $\gamma=1/4$, giving
\begin{equation}
\label{deltarhobis}
\delta\rho=\frac{1}{3}\rho_0\left(-3+\frac{2^{1-\delta}\times3^\delta\hspace{0.3mm}\delta}{\left(2-\delta\right)G^{2(\delta-1)}}\,\rho_0^{1-\delta}
\right).
\end{equation}
Similarly, we get for the pressure fluctuations
\begin{equation}
\label{deltap}
\delta p=\rho_0\left\{-w+\left(\frac{2}{3}\right)^{1-\delta}\,\frac{\delta\left[1+w\left(2-\delta\right)-\delta\right]}{\left(2-\delta\right)G^{2(\delta-1)}}\,\rho_0^{1-\delta}\right\}\,,
\end{equation}
which becomes for the radiation dominated era ($w=1/3$)
\begin{equation}
\label{raddomera}
\delta p=\frac{1}{9}\rho_0\left[-3+\frac{2^{1-\delta}\times3^\delta\,\delta\left(5-4\delta\right)}{\left(2-\delta\right)G^{2(\delta-1)}}\rho_0^{1-\delta}
\right].
\end{equation}
Notice that in the above treatment we have included
all the $\delta$-dependent terms
in $\delta\rho$ and $\delta p$. 
We stress that both these quantities vanish for $\delta\rightarrow1$, consistently with the recovery of thermal equilibrium in the radiation dominated
era within standard Cosmology.

To compute Tsallis-corrected derivative of the Ricci scalar $\mathcal{\dot{R}}$,
the trace of Einstein equation is derived by
\begin{equation}
\label{r}
\mathcal{R}=-8\pi\hspace{0.2mm}G\hspace{0.2mm} T_g\,,
\end{equation}
where
\begin{equation}
T_g=\rho-3p
\end{equation}
is the trace of the energy-momentum stress tensor. 
Using Eqs.~\eqref{deltarhobis} and~\eqref{deltap}, 
it follows that
\begin{equation}
\label{rsimple}
\mathcal{R}= \frac{2^{6-\delta}\times3^{\delta-1}\hspace{0.2mm}\pi\,\delta\left(1-\delta\right)}{\left(2-\delta\right)G^{2\delta-3}}\,\rho_0^{2-\delta}\,,
\end{equation}
where we have omitted for simplicity time-dependence. 
Therefore, the time derivative of the Ricci scalar turns out to be
\begin{equation}
\label{rdot}
\mathcal{\dot{R}}=\frac{2^{6-\delta}\times3^{\delta-1}\hspace{0.2mm}\pi\,\delta\left(1-\delta\right)}{G^{2\delta-3}}\,\rho_0^{1-\delta}\,\dot\rho_0\,.
\end{equation}
This can be further manipulated by resorting to the continuity 
equation~\eqref{ce}, here rewritten as
\begin{equation}
\dot\rho_0=-3H(1+w)\rho_0=-4H\rho_0\,,
\end{equation}
to give
\begin{equation}
\label{rdotbis}
\mathcal{\dot{R}}=\frac{2^{19/2-\delta}\times3^{\delta-3/2}\,\pi^{3/2}\,\delta\left(\delta-1\right)}{G^{2\delta-7/2}}\,\rho_0^{5/2-\delta}\,. 
\end{equation}
Following~\cite{Lamb}, here we have used the equilibrium  
form of the second Friedmann equation and of the continuity equation, since
corrections to the latter would contribute
to orders higher than that considered in the present analysis (see below). 

Equation~\eqref{rdotbis} indicates that non-extensive
Tsallis prescription can induce a non-trivial deviation from thermality
and generate baryon asymmetry, in compliance with the last 
Sakharov condition~\cite{Saka}. 

Let us now substitute the time derivative of the Ricci scalar~\eqref{rdotbis}
into the baryon asymmetry formula~\eqref{quant}. A direct calculation
leads to
\begin{equation}
\label{etabis}
\eta=\frac{35\times 2^{15/2-\delta}\times3^{\delta-1/2}\,\delta\left|\delta -1 \right|}
{\pi^{1/2}}\,\frac{g_b}{g_{*s}\,M_*^2\,T_D}\,\frac{\rho_0^{5/2-\delta}}{G^{2\delta-7/2}}\,.
\end{equation}
We express the equilibrium mass density $\rho_0$
in terms of the temperature as~\cite{Kolb}
\begin{equation}
\label{eqende}
\rho_0(T)=\frac{\pi^2\,g_{*s}}{30}\,T^4\,,
\end{equation}
to obtain 
\begin{eqnarray}
\label{neweta}
\nonumber
\eta&=&{1792\times3^{2\delta-3}\times5^{\delta-3/2}\times{\pi^{11/2-2\delta}}\,\delta\left|\delta -1 \right|}\\[2mm]
&&\times\,\frac{g_b}{g_{*}^{\delta-3/2}}
\left(\frac{T_D}{m_p}\right)^{9-4\delta}\,.
\end{eqnarray}
Here we have used $g_{*s}\approx g_{*}$, 
as discussed in the previous Section. 

Since departure of $\delta$ from unity is expected to be small (see constraints in Table~\ref{Tab1}), 
we can reasonably expand $\eta$ for $|\delta-1|\ll1$, yielding to the leading order
\begin{equation}
\label{expandetadelta}
\eta\simeq\frac{1792\left|\delta -1 \right|}{3\times5^{1/2}\,\pi^{-7/2}}\,\frac{g_b}{g_*^{-1/2}}\,\left(\frac{M_I}{m_p}\right)^5\,.
\end{equation} 
In order to estimate Tsallis parameter, we evaluate this expression at the decoupling temperature
$T_D=M_I\simeq3.3\times10^{16}\,\mathrm{GeV}$, which is the upper bound on tensor mode fluctuations at inflation\footnote{The upper bound on tensor mode fluctuations constrains the inflationary scale to be $M_I\le 3.3\times10^{16}\,\mathrm{GeV}$~\cite{Davoudiasl}.
In this setting, one has $T_D\lesssim T_{RD}\lesssim M_I$, where
$T_{RD}$ is the temperature of the radiation dominated era.
Baryon asymmetry can be large enough even for
$M_*\simeq m_p$ if $T_D\simeq M_I$ (which is the case
we are considering here). Such a scenario predicts
that tensor mode fluctuations could be soon observed~\cite{Davoudiasl}.}~\cite{Davoudiasl}.
By plugging numerical values, we  get
\begin{equation}
\label{etafin}
\eta\approx 2.38\times10^{-8}\left|\delta -1 \right|\,. 
\end{equation}

We can now constrain Tsallis exponent
by comparing the value of $\eta$ predicted by our model with the measured baryon asymmetry~\cite{Lamb,Estimate1,Estimate2,Estimate3,Estimate4,Estimate5}
\begin{equation}
\label{mba}
5.7\times10^{-11}\lesssim\eta\lesssim9.9\times10^{-11}\,.
\end{equation}
\begin{figure}[tpb]
\includegraphics[width=0.5\textwidth,trim=1 1 1 1,clip]{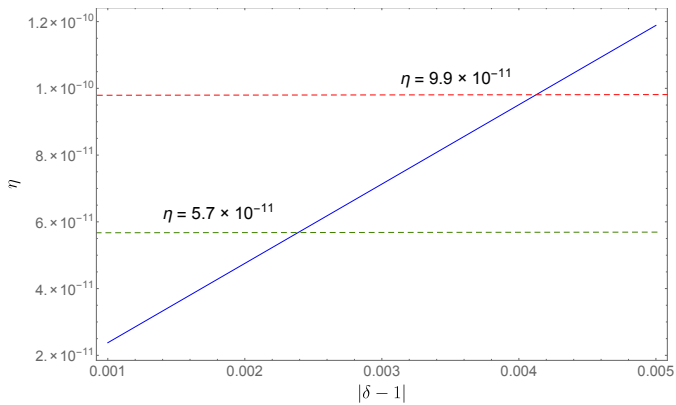}
\caption{Plot of baryon asymmetry parameter as a function of Tsallis parameter (blue curve). The red (green) dashed line represents the experimental
upper (lower) bound on $\eta$ from Eq.~\eqref{mba}.
}
\label{plot}
\end{figure}
From Eq.~\eqref{etafin}, we obtain up to $\mathcal{O}(10^{-3})$ (see Fig.~\ref{plot})
\begin{equation}
\label{finbou}
0.002\lesssim\left|\delta -1 \right|\lesssim 0.004\,.
\end{equation}
One might argue that this interval excludes the
Bekenstein-Hawking limit $\delta\rightarrow1$. The reason
for that is because in our model we are assuming that
baryogenesis is only due to corrections induced by Tsallis entropy,
while keeping the coupling between the Ricci scalar
and baryon current in its standard form. In this setting, the condition $\delta=1$ implies $\eta=0$ (see Eq.~\eqref{expandetadelta}), 
which is however in contrast with the experimental constraint~\eqref{mba}. 
On the other hand, the lower bound
on $\delta-1$ disappears if we consider  more elaborated cosmological models, in which baryon asymmetry can also be
produced by other mechanisms presented in the literature~\cite{Buras1,Georgi,Dimopoulos,Langacker,Kolb}, 
not associated to Tsallis holographic entropy itself.

In an equal fashion, Eq.~\eqref{finbou} reads as
\begin{equation}
\label{bou2}
|\delta -1 |\simeq10^{-3}\,.
\end{equation}
A posteriori, this corroborates the expansion in Eq.~\eqref{expandetadelta}.
The obtained result is in agreement with other bounds
recently obtained in cosmological scenarios and, in particular, in the study of primordial abundances of ${}^4He$, ${}^2H$ and ${}^7Li$ (see Table~\ref{Tab1}).
On the other hand, it
does not overlap with the constraint $\delta\lesssim0.5$ ($\delta<0.5$)
needed to explain the asymptotic flatness of galactic rotation curves (accelerated expansion of the present Universe) without resorting to
dark matter (dark energy)~\cite{She3}. 
Such an apparent inconsistency could be explained
by invoking some mechanism occurred during the
evolution of the Universe that may have 
reduced Tsallis parameter. This finds support in the recent result of~\cite{App13}, 
where a modified Tsallis Cosmology with a 
running exponent has been proposed to
trace the thermal history of the Universe. 
Arguments in this direction have also been
provided in Ref.~\cite{Lucianomix,Lucianomix2} 
in the context of  neutrino mixing in quantum field theory.

\begin{table}[t]
  \centering
    \begin{tabular}{|c| c c|}
    \hline
 Bound &Physical framework & Ref. \\
        \hline
                    $\delta\lesssim0.5$ & \hspace{-1.2cm} Galactic rotation curves & \cite{She3} \\[1.5mm]   
                                    $\delta<0.5$ & \hspace{0.15cm} Late-time accelerated expansion & \cite{She3} \\[1.5mm]   
         Running & \hspace{-2.95cm} Boson mixing & \cite{Lucianomix} \\[1.5mm]
$0.993\lesssim\delta\lesssim0.994$ & \hspace{-2.65cm} ${}^7Li$ Abundance & \cite{goshal} \\[1.5mm]
$\delta=1.222$ & \hspace{-1.15cm} Cosmic ray observations & \cite{Beck1} \\[1.5mm]
$0.697\le\delta\le4.2$ & \hspace{-0.4cm} Entanglement measurements & \cite{Entang} \\[1.5mm]
$|\delta-1|\lesssim 10^{-13}$ & \hspace{-0.55cm} Non-commutative geometry & \cite{Abreu} \\[1.5mm]
$\delta\ge1.203$ & \hspace{-2.15cm} Quark coalescence & \cite{Biro} \\[1.5mm]
$1\le\delta<1.333$ & \hspace{-1.5cm} High-energy collisions & \cite{HEC} \\[1.5mm]
$\delta\ge1.218 $& \hspace{-3.25cm} Black holes & \cite{Mej} \\[1.5mm]
Running & \hspace{-2.45cm} Fermion mixing  & \cite{Lucianomix2} \\[1.5mm]
$\delta\lesssim1.222$ & \hspace{-3cm} Unruh effect & \cite{LucianoGUP} \\[2mm]
$\delta\simeq 1.01$ & \hspace{-0.75cm} Big Bang Nucleosynthesis & \cite{goshal} \\[1.5mm]
$0.996\lesssim\delta\lesssim1.001$ & \hspace{-1.6cm} ${}^4He$, ${}^2H$  Abundance & \cite{goshal}\\[1.5mm]
$\delta\gtrsim0.999$ & \hspace{-0.85cm} Dark Matter relic density & \cite{goshal}\\[1.5mm]
           \hline
    \end{tabular}
  \caption{Recent bounds on Tsallis parameter coming from Cosmology, particle physics and entanglement measurements.}
  \label{Tab1}
\end{table}

Starting from the recently established connection between 
Tsallis thermostatistics and gravity-induced deformations 
of Heisenberg Uncertainty Principle (generalized uncertainty principle, GUP)~\cite{JizbaLuc}, 
in the next Section we substantiate the result~\eqref{bou2}, 
showing that it is consistent with existing bounds on the GUP parameter
predicted by many quantum gravity models.

\section{Connecting Tsallis thermostatistics and generalized uncertainty relations at Planck scale}
\label{CBT}
Several models of quantum gravity, such
as String Theory, Loop Quantum Gravity, Quantum Geometry, Doubly Special Relativity and black hole physics~\cite{Amati,Maggiore,Kempf,Scard,Capozzi,Adler,Mague} 
agree in predicting a deformation of Heisenberg
principle due to gravity effects at Planck energy. The ensuing
relation entails a minimal length and/or a maximum momentum, 
consistently with the would-be discrete structure
of spacetime arising at Planck scale.

The most common form adopted for GUP reads
\begin{equation}
\label{GUP}
\Delta x\,\Delta p \ge \frac{1}{2}\left[1+\beta\left(\frac{\Delta p}{m_p}\right)^2 \right]. 
\end{equation}
The (dimensionless) deformation parameter $\beta$
is not fixed by the theory, although investigation along
this line is active both at theoretical~\cite{Brau,Das,Pedram,ScardCas,QC,Petroz,CGgrav,BossoLuc,Gine2020} and experimental level~\cite{Bruk,GravBar,Bawaj,BossoLigo,Pendu,LucLuc} (see~\cite{Hosse} for a review).
Quite debated is also the issue of the sign of $\beta$, with 
arguments in favor of either positive~\cite{Amati,Kempf,Scard,Adler,QC,Petroz} or negative~\cite{Mague,JizbaKl,Ong,CGgrav} deformation parameter.

The emergence of a minimal length predicted by the GUP~\eqref{GUP}
should somehow affect the phase-space structure 
by modifying the elementary cell volume occupied by each quantum state. In turn, this has implications on the statistical (microscopic)
properties of quantum systems. Particularly, in~\cite{Shaba}
it has been observed that generalized statistics with a quadratic correction over Gaussian profile naturally arise as a consequence of Eq.~\eqref{GUP},  
under the condition that the total phase-space volume is kept invariant. Connections between departure from Gaussianity and
GUP have been later explored within the specific framework of Tsallis statistics~\cite{LucianoGUP}, considering the Unruh effect as a playground. 
In that case, it has been shown that modifications to
Unruh temperature induced by GUP~\cite{LucianoCas}
can be rephrased in terms of a generalized vacuum distribution, 
provided that the underlying statistics is switched
from Boltzmann-Gibbs to Tsallis prescription.

Tsallis-GUP correspondence has been rigorously formalized in~\cite{JizbaLuc}
through the study of coherent states for generalized uncertainty relations.
As a result, it has been argued that the probability distribution associated with the coherent states for the quadratic GUP~\eqref{GUP} is a Tsallis-like distribution,
the non-extensivity parameter being monotonically related to $\beta$
by
\begin{equation}
\label{GUPTsa}
\delta=\frac{\beta\hspace{0.5mm} \gamma_{\Delta p}}{m_p^2+\beta\hspace{0.5mm} \gamma_{\Delta p}}\,+\,1\,.
\end{equation}
Here, the scale-dependent parameter $\gamma_{\Delta p}$
is defined as
\begin{equation}
\label{gabe}
\gamma_{\Delta p}=\frac{2 \left(\Delta p\right)^2_\psi}{1+\beta \left(\Delta p\right)^2_\psi/m_p^2}\,,
\end{equation}
where $\left(\Delta p\right)_\psi$ is the momentum-fluctuation scale
of the system being in the coherent state $\psi$.
From the relativistic equipartition theorem, one can show that $\left(\Delta p\right)^2_\psi\simeq 12\, T^2$~\cite{JizbaLuc}. 

We can now insert Eq.~\eqref{gabe} into~\eqref{GUPTsa} and solve
for $\beta$, obtaining 
\begin{equation}
\label{49}
\beta=\frac{\left(\delta-1\right)}{12\left(5-3\delta\right)}\left(\frac{m_p}{T}\right)^2\,.
\end{equation}
By setting the temperature $T=T_D\simeq M_I$ as in the previous computation 
and using the estimate~\eqref{bou2}
for Tsallis parameter, this yields
\begin{equation}
\label{bbeta}
|\beta|\sim\mathcal{O}(10)\,.
\end{equation}
We notice that this value of the GUP parameter is 
consistent with the result of~\cite{QC}, $\beta=82\pi/5$, 
derived from quantum corrections to the Newtonian potential.
Likewise, it fits with predictions from 
Caianiello's theory of maximal acceleration~\cite{Petroz}, $\beta=8\pi^2/9$, 
and non-commutative geometry in Schwarzschild spacetime~\cite{Kana}, 
$\beta=7\pi^2/2$. Remarkably, our constraint 
is more stringent than those obtained in
other cosmological/astrophysical contexts (see~\cite{Giardino,Giardino2} for a detailed review). 
In this sense, it is much closer to predictions of string theory,
$\beta\sim\mathcal{O}(1)$, which set quantum gravity corrections
at Planck scale (see Eq.~\eqref{GUP}). 

\section{Discussion and Conclusions}
\label{DC}
The baryon asymmetry produced in the radiation dominated
era of the Universe can be explained under the assumption
of a mechanism satisfying the three Sakharov conditions. 
In this work we have speculated on a possible
Tsallis-induced asymmetry. In our picture, $CP$ is
violated by the coupling between the derivative of the Ricci scalar and the baryon current. Taking into account interactions which also
break the baryon number $B$, departure from
thermal equilibrium is generated by non-extensive Tsallis entropy.
Corrections are quantified by computing mass density and pressure fluctuations
via the modified Friedmann equations~\eqref{firsterEq} and~\eqref{secterEq}. 
These variations enter the expression
of the baryon asymmetry parameter $\eta$ through its
dependence on the time derivative of the 
Ricci scalar, which is found to be $\delta$-dependent and non-vanishing
(as opposed to the standard Boltzmann-like description). 
Requiring consistency with observed baryon asymmetry, we have constrained Tsallis parameter to be $|\delta -1 |\simeq10^{-3}$. This is in line with other bounds recently appeared in literature in Cosmology. 
Based on the connection between non-extensive Tsallis statistics
and gravitational generalizations of Heisenberg relation, we have
then transferred this result to the GUP deforming parameter, 
yielding $\beta\sim\mathcal{O}(10)$. Once more, 
this is in good agreement with many quantum gravity models
and very close to string theory predictions. Furthermore, it
is tighter than other constraints
derived in cosmological/astrophysical contexts. 
We expect that the result
here achieved can contribute to fix the ongoing discussion on the most
reliable scenario among cosmological models founded on Tsallis model.  

Apart from its own interest, the above formalism 
can find non-trivial applications in other cosmological problems. 
For instance, a demanding 
perspective is to look for signatures of tensor perturbations originated at 
inflation and propagated during Tsallis cosmological era 
in current/future experiments on the detection of primordial gravitational waves. 

There exist other extended statistic besides Tsallis proposal.
For example, a largely used statistics is that introduced by
Kaniadakis, which comes from relativistic
corrections to the Boltzmann theory~\cite{Kanad}. 
It would be interesting to explore how modified Friedmann equations
affect the mass density/pressure content of the Universe
in that context~\cite{Kanad2}. Along this line, preliminary considerations 
could be derived by making use of the relation between
Tsallis and Kaniadakis entropic measures~\cite{spara}.
Work along these directions is in progress
and will be faced elsewhere.

\acknowledgments
The author is grateful to Constantino Tsallis (Centro Brasileiro de Pesquisas Fisicas, Brazil) for useful discussion.

\end{document}